\documentclass[runningheads]{llncs}
\usepackage[T1]{fontenc}

\usepackage{pifont}
\usepackage{multirow}
\usepackage{siunitx}
\usepackage{nicefrac}
\usepackage{amsmath}
\usepackage{amsfonts}
\usepackage{bbding}
\usepackage{rotating}
\usepackage{caption}
\usepackage{subcaption}
\usepackage{cite}
\usepackage{hyperref}
\usepackage{url}
\usepackage{breakurl}
\hypersetup{breaklinks=true, colorlinks=true, pdfusetitle=true}
\usepackage{cleveref}

\usepackage{color}

\begin{document}

\title{Are Representation Disentanglement and Interpretability Linked in Recommendation Models?}
\subtitle{A Critical Review and Reproducibility Study\thanks{This preprint has not undergone peer review (when applicable) or any post-submission improvements or corrections.}}
\titlerunning{Are Disentanglement and Interpretability Linked in RSs?}

\author{
Ervin Dervishaj \inst{1} \orcidID{0000-0002-4192-1055} \and
Tuukka Ruotsalo \inst{1,2} \orcidID{0000-0002-2203-4928} \and
Maria Maistro \inst{1} \orcidID{0000-0002-7001-4817} \and
Christina Lioma \inst{1} \orcidID{0000-0003-2600-2701}
}

\authorrunning{Dervishaj et al.}

\institute{University of Copenhagen \and LUT University \\
\email{\{erde,tr,mm,c.lioma\}@di.ku.dk}}

\maketitle              
\begin{abstract}
Unsupervised learning of disentangled representations has been closely tied to enhancing the representation intepretability of Recommender Systems (RSs). This has been achieved by making the representation of individual features more distinctly separated, so that it is easier to attribute the contribution of features to the model’s predictions. However, such advantages in interpretability and feature attribution have mainly been explored qualitatively. Moreover, the effect of disentanglement on the model's recommendation performance has been largely overlooked. In this work, we reproduce the recommendation performance, representation disentanglement and representation interpretability of five well-known recommendation models on four RS datasets. We quantify disentanglement and investigate the link of disentanglement with recommendation effectiveness and representation interpretability. While several existing work in RSs have proposed disentangled representations as a gateway to improved effectiveness and interpretability, our findings show that disentanglement is not necessarily related to effectiveness but is closely related to representation interpretability. Our code and results are publicly available at \url{https://github.com/edervishaj/disentanglement-interpretability-recsys}.

\end{abstract}

\section{Introduction} \label{sec:intro}
In order to provide users with a personalised experience, recommender systems (RSs) build representations that encode user preferences from behavioural data. One way of extracting informative and more interpretable representations is by learning disentangled representations~\cite{schmidhuber1992learning,bengio2013representation,chen2016infogan,lake2017building,locatello2019challenging,wang2022disentangled}. Real-world RSs data contain complex combinations of user preference factors (e.g., in a movie RS: movie genre, movie actors, mood of the user, time of the day, etc.). Given that one factor is often invariant to changes in other factors, unsupervised disentangled representation learning aims to encode each factor onto one of the dimensions of the latent space.
Existing work claim that such representations improve interpretability because each latent dimension captures one semantically-meaningful explanatory factor of the data~\cite{bengio2013representation,chen2016infogan,higgins2016beta,kim2018disentangling,chen2018isolating,chen2016infogan,burgess2018understanding,dupont2018learning,locatello2019challenging,wang2022disentangled} (also called \emph{factor of variation}).
Disentangled representations have been recently studied -- in RSs, and more generally -- under the class of deep generative models, namely variational autoencoders (VAEs)~\cite{kingma2013auto} and generative adversarial networks (GANs)~\cite{goodfellow2014generative}. In order to \emph{force} the deep neural network to disentangle the factors of variation of the data in the latent space, previous work modify the model's vanilla objective function with a regularization term~\cite{locatello2019challenging}, thereby acting as an \emph{``interpretability constraint on the latent dimensions''}~\cite{rudin2022interpretable}. This change in the objective function has been empirically linked to reduced reconstruction capabilities in VAEs~\cite{higgins2016beta}, pointing to a trade-off between disentanglement and downstream task performance~\cite{burgess2018understanding,kim2018disentangling,lezama2019overcoming}.

Recently, the RS community has explored unsupervised disentangled representations for more accurate modeling of user preferences. However, previous work provides only a qualitative perspective on the model disentanglement, usually in the form of a visual inspection of the learned representations reduced to 2 dimensions~\cite{ma2019learning,wang2020disentangled,zheng2021disentangling,guo2022topicvae,de2023disentangling}.
The lack of an objective evaluation of disentanglement makes it unclear whether the findings of disentanglement, model effectiveness and improved representation interpretability generalize over different models and datasets.

In this work, we seek to address the following research questions:
\begin{enumerate}
    \item Can we reproduce the recommendation effectiveness and disentanglement of existing RSs models aimed at learning disentangled representations?
    \item What is the effect of disentanglement on recommendation effectiveness?
    \item What is the effect of disentanglement on RSs representation interpretability?
\end{enumerate}
\noindent First, we focus on reproducing recommendation effectiveness and disentanglement of state-of-the-art unsupervised disentangled RSs. Our results show that the reproducibility of the reported effectiveness of these models is dataset dependent, with differences up to 43\% across datasets. To our best knowledge, only~\cite{nema2021disentangling} has utilized disentanglement metrics to quantify disentanglement in RSs. However, we failed to reproduce their results which prompted us to conduct an empirical investigation that uses existing disentanglement metrics (\emph{disentanglement} and \emph{completeness}~\cite{eastwood2018framework}). Specifically, we measure the disentanglement of user representations of five recommendation models on four datasets. 
Second, we provide the first study that quantifies representation interpretability with feature attribution methods and links it to disentanglement in RSs. We adapt two existing feature attribution approaches: LIME~\cite{ribeiro2016should} and Shapley Value (SHAP)~\cite{lundberg2017unified}. We use the feature-level scores produced by LIME and SHAP to define two measures, called LIME-global and SHAP-global, which quantify the degree of interpretability of a model's representations. Finally, through a correlation analysis, we study the link between disentanglement, representation interpretability and effectiveness in RSs. Our findings do not support the alleged trade-off between disentanglement and effectiveness~\cite{higgins2016beta,kim2018disentangling,lezama2019overcoming,rudin2022interpretable}, due to no consistent statistically significant relation between the two. However, in line with prior qualitative work in RSs on interpretability and disentanglement~\cite{ma2019learning,wang2020disentangled,nema2021disentangling}, we find representation interpretability to be positively related to disentanglement.

\section{Background}

\subsection{Disentangled Representations}
Disentangled representations aim to separate the \emph{factors of variation} in the data, i.e., changing one aspect of a data point should only affect the latent dimension responsible for the changed aspect, while keeping all the other dimensions unchanged. Since the data generating function is usually governed by few explanatory factors, separating these factors in a latent space makes for a more interpretable representation.
Recently proposed models for disentangled representations are largely based on VAEs~\cite{higgins2016beta,locatello2019challenging}. $\beta$-VAE~\cite{higgins2016beta} was first to apply VAEs to learn a factorized representation of the independent explanatory factors from the data by enforcing an information bottleneck through a new hyperparameter $\beta > 1$ which penalizes the KL-divergence term in the VAEs' evidence lower bound (ELBO) objective.

\subsection{Disentangled Representations in RSs}
Disentangled representation learning has shown good empirical results by modeling the implicit biases in the data~\cite{higgins2016beta,kim2018disentangling,chen2018isolating,burgess2018understanding,shi2019variational,meo2024alpha}. Inspired by this, the RS community has recently explored unsupervised disentangled representations: based on user intents~\cite{ma2019learning,wang2020disentangled,10.1145/3539618.3591665,zhang2024denoising}, and supervised disentangled representations: based on item topics~\cite{guo2022topicvae}, user conformity to group (i.e., popularity bias)~\cite{zheng2021disentangling,zhao2022popularity,yang2023debiased}, long and short term user interests~\cite{zheng2022disentangling} and causal components~\cite{wang2022causal,gao2023cirs}. Supervised disentangling models select some semantically meaningful item attributes, such as item topic~\cite{guo2022topicvae}, item attribute~\cite{nema2021disentangling}, etc., which are then used during training such that one dimension of the representation space encapsulates only one such item attribute.
Unsupervised disentangling models, on the other hand, have the advantage that they do not need user/item side information, often not available in RSs datasets. The state of the art in RSs is MacridVAE~\cite{ma2019learning}, which assumes that user interactions are based on several user intent so it disentangles the user preferences into macro concepts (e.g., item categories). Ma et al.~\cite{ma2019learning} represent each macro concept with a $d$-dimension vector to allow for a finer granularity of the user preferences, which they further disentangle by penalizing the KL-divergence in their VAE objective with $\beta \gg 1$ (similar to~\cite{higgins2016beta}).

\subsection{Disentanglement, Interpretability of Representations and Effectiveness}
Disentangled representations have been linked to more interpretable representations~\cite{chen2016infogan,higgins2016beta}, but at the expense of downstream task performance~\cite{burgess2018understanding,kim2018disentangling,lezama2019overcoming}. This is due to the regularization effect of disentanglement~\cite{locatello2019challenging,rudin2022interpretable} during the learning process. In RSs, several work show qualitatively that disentanglement helps in learning representations that are more interpretable;~\cite{ma2019learning} inspects the ability of disentangled representations in capturing the true item categories,~\cite{wang2020disentangled,li2021learning} investigate how closely users' item reviews relate to the disentangled user intents and~\cite{nema2021disentangling} compares the correlation in latent space between users and items on supervised disentangled dimensions and unsupervised ones, separately. While prior work on disentanglement in RSs proposes models that achieve state-of-the-art performance, the regularization aspect of disentanglement, and a possible trade-off with effectiveness, remains largely unexplored and only briefly mentioned by Nema et al. (see section 5.3 in~\cite{nema2021disentangling}).

\section{Experiments}
In this section we describe the experimental setup of our reproducibility study on unsupervised disentangled RSs and the relation between disentanglement, recommendation effectiveness and representation interpretability. We describe the effectiveness, disentanglement and our proposed representation interpretability measures, the datasets and the recommendations models that we reproduce.

\subsection{Measures}
\subsubsection{Recommendation effectiveness measures}
We use Normalized Discounted Cumulative Gain (NDCG) \cite{jarvelin2002cumulated}, recall, Mean Reciprocal Rank (MRR)~\cite{voorhees1999proceedings} and coverage \cite{ge2010beyond}, at cutoff 10, 50 and 100.

\subsubsection{Disentanglement measures}
Several measures of disentanglement have been proposed, e.g., BetaVAE~\cite{higgins2016beta}, FactorVAE~\cite{kim2018disentangling}, or the Dis\-en\-tan\-gle\-ment-Com\-ple\-te\-ness-Informativeness (DCI) framework~\cite{eastwood2018framework}. All of these measures require ground truth factors of variation (see \cref{sec:datasets} on how we recover these ground truth factors). BetaVAE and FactorVAE also require a ground truth simulation function that, given factors of variation, can produce data samples. Since in RSs we lack such a simulation function, we use two measures from the DCI framework:
\emph{disentanglement} and \emph{completeness}\footnote{We use the code release by~\cite{eastwood2018framework} and the \texttt{disentanglement\_lib} framework~\cite{locatello2019challenging}.}. Both of these measures use simple estimators to predict the ground truth factors of variation of the data from a learned disentangled space. We describe these two measures next.

We assume that a model can learn an $M$-dimensional latent representation $z \in \mathbb{R}^{M}$. Given $K$ binary ground truth factors of variation, we train $K$ binary classifiers $f_{j}: \mathbb{R}^{M} \rightarrow \{0,1\}$ for $j \in 1 \ldots K$, that, given the latent representation $z$, predict the presence of each of the $K$ factors. We collect in a matrix $F \in \mathbb{R}^{M \times K}$ the importance\footnote{We use \texttt{GradientBoostingClassifier} from \texttt{scikit-learn} package as our binary classifier which provides impurity-based feature importances.} of dimension $i$ of the latent space in predicting factor $j$. Then \emph{disentanglement} ($\mathbf{D}$) and \emph{completeness} ($\mathbf{C}$) are computed as:
\begin{equation}
    \begin{aligned}[c]
        &\mathbf{D} = \sum^{M}_{i=1} \alpha_{i} D_{i} \,, \: D_{i} = 1 - H_{K}(P_{i}) \\
        &\alpha_{i} = \nicefrac{\sum^{K}_{j=1} F_{ij}}{\sum^{K}_{j=1} \sum^{M}_{i=1} F_{ij}} \\
        &H_{K}(P_{i}) = -\sum^{K}_{j=1} P_{ij} \log_{K} P_{ij} \\
        & P_{ij} = \nicefrac{F_{ij}}{\sum^{K}_{j=1} F_{ij}}
    \end{aligned}
    \qquad
    \begin{aligned}[c]
        &\mathbf{C} = \sum^{K}_{j=1} \beta_{j} C_{j} \,, \: C_{j} = 1 - H_{M}(\Tilde{P}_{j}) \\
        &\beta_{j} = \nicefrac{\sum^{M}_{i=1} F_{ij}}{\sum^{K}_{j=1} \sum^{M}_{i=1} F_{ij}} \\
        &H_{M}(\Tilde{P}_{j}) = -\sum^{M}_{i=1} \Tilde{P}_{ij} \log_{M} \Tilde{P}_{ij} \\
        &\Tilde{P}_{ij} = \nicefrac{F_{ij}}{\sum^{M}_{i=1} F_{ij}}
    \end{aligned}
\end{equation}
$\mathbf{D}$ and $\mathbf{C}$ are expressed as weighted sums of the \emph{disentanglement} $D_i$ of each dimension in the learned representation, and of the \emph{completeness} $C_j$ of each factor of variation, respectively. The entropy $H$ specifies how feature importance probabilities $P_j$ and $\Tilde{P}_j$ are distributed across the factors of variation and the dimensions of the representations, respectively. A higher entropy means that the classifier's feature importance is more uniformly distributed across factors/dimensions resulting in lower \emph{disentanglement}/\emph{completeness}, whereas a lower entropy means placing higher feature importance on a smaller subset of the dimensions of the learned representations, thus achieving better overall disentanglement. Both $\mathbf{D}$ and $\mathbf{C}$ range in $[0,1]$, and the higher, the more disentangled/complete the representations. In this work, we use gradient boosting decision trees as binary classifiers and tune them on the representations of users in the validation set (\cref{sec:datasets}).

\subsubsection{Interpretability measures} \label{subsec:interpretability}
LIME~\cite{ribeiro2016should}, SHAP~\cite{lundberg2017unified} and Integrated Gradients (IG)~\cite{sundararajan2017axiomatic} are well-known local interpretability methods, i.e., they measure feature importance on the model's prediction per data sample. However, to our best knowledge, there exists no method that quantifies representation interpretability. To quantify the interpretability of the disentangled representations, we utilise the $K$ binary classifiers used for \emph{disentanglement} and \emph{completeness} and adapt LIME and SHAP\footnote{We do not use IG because the gradient boosting trees classifiers are non-differentiable.} into global measures, called LIME-global and SHAP-global. For each classifier $f_j$, we collect the mean absolute LIME/SHAP latent dimension importance across all user representations into a column vector $\mathbf{s}_j \in \mathbb{R}^M$. We concatenate all vectors $\mathbf{s}_j$ to build matrix $\mathcal{S} \in \mathbb{R}^{M \times K}$. As a final value for LIME-/SHAP-global, we normalize the columns of $\mathcal{S}$ into $[0,1]$ and take the mean of the Jensen-Shannon (JS) divergence computed between every pair of columns of $\mathcal{S}$. The intuition behind LIME-/SHAP-global and the application of the JS divergence is to penalise redundancy in the dimensions of the representation space and to promote sparsity in feature importance. If two binary classifiers place similar feature importance on the same subset of dimensions of the learned representation space, then the divergence between the classifiers' feature importance distribution will be 0. In this way, LIME-/SHAP-global penalises redundancy in the learned representations. Moreover, for the JS divergence (i.e., interpretability) to increase, the feature importance distribution should be concentrated only on some features (i.e., promoting sparsity in the features, and resulting in a simpler and more interpretable model~\cite{bohanec1994trading,doshi2017towards,guidotti2018survey,rudin2022interpretable}). LIME-/SHAP-global range is $[0,1]$, with higher values denoting more interpretable representations.

\begin{table}[htbp]
    \centering
    \caption{Dataset statistics after preprocessing. Min. IPU/IPI are the \emph{minimum interactions per user/item}.}
    \begin{tabular}{c|r|r|r|r}
        \hline
        & Amazon-CD & ML1M & Yelp & GR-Children \\
        \hline
        Interactions & \num{570747} & \num{1000209} & \num{2762098} & \num{5067546}\\
        Users & \num{23024} & \num{6040} & \num{99011} & \num{117293}\\
        Items & \num{19444} & \num{3706} & \num{56441} & \num{42119}\\
        Min. IPU & \num{10} & \num{20} & \num{10} & \num{10}\\
        Min. IPI & \num{10} & \num{1} & \num{10} & \num{10}\\
        Sparsity & \qty{99.877}{\percent} & \qty{95.532}{\percent} & \qty{99.953}{\percent} & \qty{99.897}{\percent}\\
        \hline
    \end{tabular}
    \label{tab:dataset_stats}
\end{table}

\begin{table}[htbp]
    \centering
    \caption{Model summary. \emph{NP} = non-personalised, \emph{MF} = matrix factorisation, \emph{DAE} = denoising autoencoder, \emph{VAE} = variational autoencoder, \emph{DIS} = explicitly disentangling model.}
    \begin{tabular}{c|c|c|c|c|c}
        \hline
        \multirow{2}{*}{Model} & \multicolumn{5}{c}{Model Type} \\
        & NP & MF & DAE & VAE & DIS \\
        \hline
        Top-Popular & \ding{51} & & & & \\
        PureSVD & & \ding{51} & & & \\
        MultiDAE & & & \ding{51} & & \\
        MultiVAE & & & & \ding{51} & \\
        $\beta$-VAE & & & & \ding{51} & \ding{51} \\
        MacridVAE & & & & \ding{51} & \ding{51} \\
        \hline
    \end{tabular}
    \label{tab:models}
\end{table}

\subsection{Datasets} \label{sec:datasets}
\Cref{tab:dataset_stats} shows the statistics of the datasets. We reproduce results for MovieLens 1M\footnote{\url{https://grouplens.org/datasets/movielens}}~\cite{harper2015movielens} and GoodReads-Children\footnote{\url{https://mengtingwan.github.io/data/goodreads.html\#datasets}}~\cite{wan2018item,wan2019fine}, both used by unsupervised disentangled models in RSs. In addition, we evaluate our reproduced models also on two other datasets -- Amazon-CD\footnote{\url{https://cseweb.ucsd.edu/~jmcauley/datasets/amazon_v2}}~\cite{ni2019justifying} and Yelp\footnote{\url{https://www.yelp.com/dataset}} -- that were not in the original papers, in order to investigate the generalizability of our results. We focus on user-item interactions, so we binarize the ratings by setting to $1$ all ratings~$\geq 1$ and everything else to $0$. 

\subsubsection{Sampling and splits}
We use $10$-core~\cite{he2017neural,nema2021disentangling} sampling\footnote{We use the \texttt{recpack}~\cite{michiels2022recpack} Python package to handle filtering, sampling and splitting of the datasets.} for all datasets except for ML1M, which includes only users with at least 20 interactions. Since we evaluate models that build user representations, a timestamp split might exclude some of the users during training. For this reason, we use random per-user train-validation-test splits with a ratio of 3:1:1. 

\subsubsection{Ground truth factors}
Quantifying disentanglement requires access to the ground truth factors of variation of the data~\cite{eastwood2018framework,locatello2019challenging,higgins2016beta,kim2018disentangling}. To our best knowledge, no publicly available RSs dataset provides such factors. In this reproducibility study we follow closely~\cite{nema2021disentangling} and utilize item content information to compose ground truth factors. Amazon-CD, ML1M, and Yelp provide tags/categories for each item. We keep the  100 most popular tags/categories, which we then group in 20 clusters using k-means clustering, with each tag/category represented as a vector over the items. Each item is part of a subset of clusters based on its corresponding tags/categories\footnote{In ML1M, each tag is represented as a relevance score vector over the items. An item is assigned a cluster if the average relevance score of its tags that fall in the cluster is greater than $M$. We set $M=0.4$ as per~\cite{nema2021disentangling}.}. Finally, a user is assigned to a cluster if at least 50\% of the items that the user has interacted with are part of that cluster. GoodReads-Children provides bookshelves for each book. We first order the bookshelves in decreasing order according to the number of books they contain and then manually merge bookshelves with similar names (e.g., `picturebooks' and `picture-book') and drop bookshelves whose name is not semantically meaningful (e.g., `to-read', `books-i-own')\footnote{We follow~\cite{nema2021disentangling}, where uninformative bookshelves are dropped if marked as such by all authors.} to get a short final list of bookshelves. Finally, we assign users to shelves if at least 50\% of their rated books fall within a shelf. We consider the clusters of categories (Amazon-CD, ML1M, Yelp) and the bookshelves (GoodReads-Children) as each dataset's respective set of ground truth factors of variation.

\subsection{Recommendation Models} \label{sec:models}
We focus on unsupervised disentangling models given the lack of ground truth factors of variation in RSs. \Cref{tab:models} lists our models. We reproduce two unsupervised disentangling models: MacridVAE~\cite{ma2019learning} and $\beta$-VAE\footnote{We adapt $\beta$-VAE to use the multinomial distribution, similar to MultiVAE, and we set $\beta > 1$ during tuning to enforce disentanglement~\cite{higgins2016beta}.}~\cite{higgins2016beta} and four non-disentangling models, as additional baselines: Top-Popular recommends only the most popular items, PureSVD~\cite{cremonesi2010performance} is a simple matrix factorization (MF) model, MultiDAE~\cite{liang2018variational} and MultiVAE~\cite{liang2018variational} are AE-based RSs. For each model, we use the official implementation released by the authors. We use the \texttt{hyperopt}~\cite{bergstra2013making} Python package to tune hyperparameters through 50 runs of Bayesian search optimization. All models are tuned for NDCG@100. For a fair comparison, we set common hyperparameter ranges -- log-uniform distribution in $[\exp(-10),\exp(-2)]$ for \emph{learning rate}, integer uniform distribution in $[2,20]$ for \emph{latent dimensionality}, $\{128, 256, 512, 1024\}$ for \emph{batch size} -- and constrain training to a maximum of 500 epochs with early stopping. In selecting the hyperparameter tuning ranges\footnote{We provide the optimal hyperparameters with our codes.}, we make sure to include all the extremes of the ranges reported by the authors of the models that we reproduce. We use five different randomization seeds for both model initialization and dataset splitting, and report mean scores of the models from the five seeds on the unseen test set.

\begin{table}[htbp]
    \scriptsize
    \centering
    \caption{Effectiveness and disentanglement reproducibility results for~\cite{nema2021disentangling} and~\cite{ma2019learning}. Their results are shown as reported in the papers. Our results represent the mean of 5 runs tuned according to \cref{sec:models}.}
    \begin{tabular}{c|c|c|c|c|c|c|c|c}
        \hline
        \multirow{2}{*}{Model} & \multicolumn{4}{c|}{ML1M} & \multicolumn{4}{c}{GR-CHILDREN} \\
        & N@100 & R@50 & D & C & N@100 & R@50 & D & C \\
        \hline
        MultiDAE~\cite{ma2019learning} & 0.4045 & 0.4678 & - & - & - & - & - & - \\
        MultiDAE~\cite{nema2021disentangling} & 0.4040 & 0.4670 & 0.3810 & 0.3120 & 0.4250 & 0.5910 & 0.2870 & 0.2430 \\
        MultiDAE (ours) & 0.4366 & 0.4500 & 0.1777 & 0.2605 & 0.3452 & 0.3807 & 0.1309 & 0.1849 \\
        \hline
        \emph{Max rel. change} & $\uparrow 8.0\%$ & $\downarrow 3.8\%$ & $\downarrow 53.4\%$ & $\downarrow 16.5\%$ & $\downarrow 18.8\%$ & $\downarrow 35.6\%$ & $\downarrow 54.4\%$ & $\downarrow 23.9\%$ \\
        \hline
        \hline
        
        MultiVAE~\cite{ma2019learning} & 0.4056 & 0.4583 & - & - & - & - & - & - \\
        MultiVAE~\cite{nema2021disentangling} & 0.4050 & 0.4580 & 0.3610 & 0.2940 & 0.4040 & 0.5770 & 0.3080 & 0.2630 \\
        MultiVAE (ours) & 0.4360 & 0.4456 & 0.1859 & 0.2894 & 0.3432 & 0.3766 & 0.1377 & 0.1805\\
        \hline
        \emph{Max rel. change} & $\uparrow 7.7\%$ & $\downarrow 2.8\%$ & $\downarrow 48.5\%$ & $\downarrow 1.6\%$ & $\downarrow 15.0\%$ & $\downarrow 34.7\%$ & $\downarrow 54.4\%$ & $\downarrow 31.4\%$ \\
        \hline
        \hline
        
        $\beta$-VAE~\cite{ma2019learning} & 0.4056 & 0.4582 & - & - & - & - & - & - \\
        $\beta$-VAE~\cite{nema2021disentangling} & 0.4540 & 0.4110 & 0.7450 & 0.4730 & 0.4150 & 0.5860 & 0.4840 & 0.3030 \\
        $\beta$-VAE (ours) & 0.4070 & 0.4164 &  0.1879 & 0.2059 & 0.2999 & 0.3335 & 0.1053 & 0.2010 \\
        \hline
        \emph{Max rel. change} & $\downarrow 10.4\%$ & $\downarrow 9.1\%$ & $\downarrow 74.8\%$ & $\downarrow 56.5\%$ & $\downarrow 27.7\%$ & $\downarrow 43.1\%$ & $\downarrow 78.2\%$ & $\downarrow 33.7\%$ \\
        \hline
        \hline
        
        MacridVAE~\cite{ma2019learning} & 0.4274 & 0.4904 & - & - & - & - & - & - \\
        MacridVAE (ours) & 0.4580 & 0.4774 & 0.1593 & 0.2527 & 0.3764 & 0.4082 & 0.1593 & 0.2527 \\
        \hline
        \emph{Max rel. change} & $\uparrow 7.2\%$ & $\downarrow 2.7\%$ & - & - & - & - & - & - \\
        \hline
    \end{tabular}
    \label{tab:reproducibility}
\end{table}

\section{Results and Discussion} \label{sec:results}
\subsection{Reproducibility Results (RQ1)} \label{sec:reproducibility}
In our literature review on disentanglement for RSs, we found~\cite{nema2021disentangling} as the only work that explicitly measures disentanglement with existing metrics, so we focus on reproducing their reported results and the results of MacridVAE~\cite{ma2019learning}. In~\cref{tab:reproducibility} we give the original and reproduced results\footnote{Note that the authors of MacridVAE do not report disentanglement scores in their experiments.}.

Regarding recommendation effectiveness, we observe some small discrepancies in ML1M; our NDCG@100 scores are generally higher (up to 8\%) than the reported ones, with the exception of $\beta$-VAE which is 10\% lower. Our recall@50 scores show a smaller difference from the original work results with $\beta$-VAE up to 9\% lower score. In GoodReads-Children, we observe much larger differences with NDCG@100 and recall@50 up to 28\% and 43\% lower than the reported scores, respectively. We attribute the discrepancies to i) how we binarize explicit ratings; both~\cite{nema2021disentangling} and~\cite{ma2019learning} set ratings $>4$ to 1, whereas we set to 1 all ratings $>1$, and ii) how we tune the hyperparameters, especially the latent space dimensionality; both~\cite{nema2021disentangling} and~\cite{ma2019learning} use a fixed latent dimensionality, meanwhile we tune it according to~\cref{sec:models}. 

Regarding disentanglement, we observe significant discrepancies in our reproduced results. Even though Nema et al.~\cite{nema2021disentangling} clearly describe the formulations of the metrics, we were not able to reproduce their scores, despite trying to implement their formulations as faithfully as possible. Moreover, retrieving the ground truth factors of variation -- used to compute disentanglement scores -- involves some hyperparameters; the threshold $M$ of average relevance score for assigning an item to a cluster of its tags in ML1M, which we set according to~\cite{nema2021disentangling}, and the merging/dropping of bookshelves in GoodReads-Children. As a final attempt, we tried using logistic regression and random forests as classifiers, but the results were still very different. After these efforts, we reached out to the authors, but the code was not made available to us.

\begin{table}[htbp]
    \scriptsize
    \centering
    \caption{Mean and standard deviation of recommendation effectiveness over 5 runs. Models tuned for NDCG@100. Best effectiveness results are given in bold.}
    \begin{tabular}{c|c|c|c|c|c}
        \hline
        \multirow{2}{*}{Dataset} & \multirow{2}{*}{Model} & \multicolumn{4}{c}{EFFECTIVENESS} \\
        & & NDCG@10 & RECALL@10 & MRR@10 & COVERAGE@10\\
        \hline
        \multirow{6}{*}{
            \begin{turn}{62}
                \centering
                Amazon-CD
            \end{turn}
        } & Top-Popular &  0.0059 ± 0.0002 &  0.0084 ± 0.0002 &  0.0098 ± 0.0005 &  0.0000 ± 0.0000 \\
        & PureSVD &  0.0288 ± 0.0003 &  0.0386 ± 0.0007 &  0.0462 ± 0.0006 &  0.0326 ± 0.0006 \\
        & MultiDAE &  0.0806 ± 0.0016 &  0.1032 ± 0.0018 &  0.1216 ± 0.0030 &  0.4761 ± 0.1953 \\
        & MultiVAE &  0.0853 ± 0.0019 &  0.1085 ± 0.0028 &  0.1295 ± 0.0028 &  0.4664 ± 0.1194 \\
        & $\beta$-VAE &  0.0570 ± 0.0090 &  0.0762 ± 0.0108 &  0.0864 ± 0.0140 &  0.4067 ± 0.1980 \\
        & MacridVAE &  \textbf{0.1621 ± 0.0039} &  \textbf{0.1824 ± 0.0029} &  \textbf{0.2560 ± 0.0076} &  \textbf{0.6624 ± 0.0181}\\
        \hline
        \multirow{6}{*}{
            \begin{turn}{62}
                \centering
                ML1M
            \end{turn}
        } & Top-Popular &  0.0533 ± 0.0030 &  0.0556 ± 0.0038 &  0.1244 ± 0.0049 & 0.0000 ± 0.0000 \\
        & PureSVD &  0.3889 ± 0.0008 &  0.3669 ± 0.0011 &  0.6201 ± 0.0006 &  0.1792 ± 0.0016 \\
        & MultiDAE &  0.3806 ± 0.0056 &  0.3632 ± 0.0042 &  0.6142 ± 0.0116 &  0.4420 ± 0.0348 \\
        & MultiVAE &  0.3853 ± 0.0047 &  0.3656 ± 0.0042 &  0.6213 ± 0.0056 &  0.4611 ± 0.0275 \\
        & $\beta$-VAE &  0.3553 ± 0.0222 &  0.3381 ± 0.0183 &  0.5822 ± 0.0287 &  0.4032 ± 0.0145 \\
        & MacridVAE &  \textbf{0.3963 ± 0.0020} &  \textbf{0.3807 ± 0.0022} &  \textbf{0.6269 ± 0.0012} &  \textbf{0.4793 ± 0.0188} \\
        \hline
        \multirow{6}{*}{
            \begin{turn}{62}
                \centering
                Yelp
            \end{turn}
        } & Top-Popular &  0.0029 ± 0.0003 &  0.0042 ± 0.0004 &  0.0051 ± 0.0007 &  0.0002 ± 0.0000 \\
        & PureSVD &  0.0304 ± 0.0002 &  0.0388 ± 0.0003 &  0.0544 ± 0.0002 &  0.0109 ± 0.0001 \\
        & MultiDAE &  0.0586 ± 0.0010 &  0.0756 ± 0.0010 &  0.1000 ± 0.0018 &  0.2567 ± 0.0514 \\
        & MultiVAE &  0.0569 ± 0.0015 &  0.0733 ± 0.0016 &  0.0973 ± 0.0031 &  0.2835 ± 0.0544 \\
        & $\beta$-VAE &  0.0458 ± 0.0051 &  0.0597 ± 0.0053 &  0.0788 ± 0.0096 &  0.2212 ± 0.0441 \\
        & MacridVAE &  \textbf{0.0650 ± 0.0013} &  \textbf{0.0833 ± 0.0016} &  \textbf{0.1102 ± 0.0021} &  \textbf{0.3387 ± 0.0150} \\
        \hline
        \multirow{6}{*}{
            \begin{turn}{62}
                \centering
                GR-Children
            \end{turn}
        } & Top-Popular &  0.0349 ± 0.0095 &  0.0452 ± 0.0100 &  0.0654 ± 0.0206 & 0.0000 ± 0.0000 \\
        & PureSVD &  0.2300 ± 0.0006 &  0.2498 ± 0.0008 &  0.3684 ± 0.0008 &  0.0053 ± 0.0001 \\
        & MultiDAE &  0.3452 ± 0.0033 &  0.3807 ± 0.0033 &  0.4989 ± 0.0040 &  0.2233 ± 0.0538 \\
        & MultiVAE &  0.3432 ± 0.0022 &  0.3766 ± 0.0034 &  0.4989 ± 0.0027 &  0.2239 ± 0.0429 \\
        & $\beta$-VAE &  0.2999 ± 0.0096 &  0.3335 ± 0.0081 &  0.4426 ± 0.0147 &  \textbf{0.3175 ± 0.0912} \\
        & MacridVAE &  \textbf{0.3764 ± 0.0038} &  \textbf{0.4082 ± 0.0036} &  \textbf{0.5406 ± 0.0046} &  0.3078 ± 0.0189 \\
        \hline
    \end{tabular}
    \label{tab:effectiveness}
\end{table}

To investigate the generalizability of these results, we evaluate the reproduced models on two additional datasets (Amazon-CD and Yelp) and include PureSVD, as a MF baseline model that learns representations from interactions, and Top-Popular. For a fair comparison, we tune all the models according to~\cref{sec:models} and train them on implicit ratings (all ratings $>1$ set to 1). We present these results in \cref{tab:effectiveness}. \Cref{tab:effectiveness} shows MacridVAE as the best model across all effectiveness measures and all datasets (in Amazon-CD twice, and in GoodReads-Children 6-8\% better than the next best model), which is consistent with~\cite{ma2019learning}. Overall, neural models -- MultiVAE and MultiDAE -- perform much better than the MF model, as reported also by~\cite{ferrari2019we}.

\begin{table}[htbp]
    \scriptsize
    \centering
    \caption{Mean and standard deviation of disentanglement and interpretability over 5 seeds. Models tuned for NDCG@100. Best disentanglement and interpretability results are given in bold.}
    \begin{tabular}{c|c|c|c|c|c}
        \hline
        \multirow{2}{*}{Dataset} & \multirow{2}{*}{Model} & \multicolumn{2}{c|}{DISENTANGLEMENT} & \multicolumn{2}{c}{INTERPRETABILITY} \\
        & & DISENTANG. & COMPLETE. & LIME-global & SHAP-global \\
        \hline
        \multirow{6}{*}{
            \begin{turn}{62}
                \centering
                Amazon-CD
            \end{turn}
        } & Top-Popular & - & - & - & -\\
        & PureSVD & {\textbf{0.2637 ± 0.0082}} & {\textbf{0.3067 ± 0.0077}} & 0.4586 ± 0.0179 & 0.5945 ± 0.0201\\
        & MultiDAE & 0.2255 ± 0.0333 & 0.2568 ± 0.0266 & 0.4285 ± 0.0177 & 0.5207 ± 0.0507\\
        & MultiVAE & 0.2144 ± 0.0079 & 0.2473 ± 0.0145 & 0.4246 ± 0.0124 & 0.5377 ± 0.0353\\
        & $\beta$-VAE & 0.1301 ± 0.0458 & 0.3054 ± 0.1153 & 0.3379 ± 0.0518 & 0.4284 ± 0.0564\\
        & MacridVAE & 0.2024 ± 0.0443 & 0.2611 ± 0.0274 & {\textbf{0.4669 ± 0.0121}} & {\textbf{0.6267 ± 0.0502}}\\
        \hline
        \multirow{6}{*}{
            \begin{turn}{62}
                \centering
                ML1M
            \end{turn}
        } & Top-Popular & - & - & - & -\\
        & PureSVD & 0.1565 ± 0.0098 & {\textbf{0.3586 ± 0.0116}} & 0.3035 ± 0.0067 & 0.4390 ± 0.0143\\
        & MultiDAE & 0.1777 ± 0.0179 & 0.2605 ± 0.0088 & 0.3351 ± 0.0051 & {\textbf{0.4784 ± 0.0297}}\\
        & MultiVAE & {\textbf{0.1859 ± 0.0277}} & 0.2894 ± 0.0202 & 0.3168 ± 0.0144 & 0.4694 ± 0.0280\\
        & $\beta$-VAE & 0.1305 ± 0.0298 & 0.3251 ± 0.0718 & 0.3013 ± 0.0346 & 0.4109 ± 0.0317\\
        & MacridVAE & 0.1207 ± 0.0091 & 0.2080 ± 0.0381 & {\textbf{0.3367 ± 0.0204}} & 0.4309 ± 0.0332\\
        \hline
        \multirow{6}{*}{
            \begin{turn}{62}
                \centering
                Yelp
            \end{turn}
        } & Top-Popular & - & - & - & -\\
        & PureSVD & {\textbf{0.2987 ± 0.0316}} & {\textbf{0.2687 ± 0.0170}} & {\textbf{0.5200 ± 0.0234}} & {\textbf{0.6601 ± 0.0296}}\\
        & MultiDAE & 0.1909 ± 0.0391 & 0.1917 ± 0.0309 & 0.4744 ± 0.0487 & 0.5330 ± 0.0549\\
        & MultiVAE & 0.2190 ± 0.0423 & 0.2323 ± 0.0235 & 0.4884 ± 0.0298 & 0.5712 ± 0.0496\\
        & $\beta$-VAE & 0.1879 ± 0.0148 & 0.2059 ± 0.0199 & 0.4733 ± 0.0179 & 0.5468 ± 0.0133\\
        & MacridVAE & 0.2208 ± 0.0817 & 0.2366 ± 0.0596 & 0.4473 ± 0.0560 & 0.5571 ± 0.1381\\
        \hline
        \multirow{6}{*}{
            \begin{turn}{62}
                \centering
                GR-Children
            \end{turn}
        } & Top-Popular & - & - & - & -\\
        & PureSVD & 0.1219 ± 0.0059 & {\textbf{0.2787 ± 0.0097}} & 0.4131 ± 0.0081 & 0.4376 ± 0.0104\\
        & MultiDAE & 0.1309 ± 0.0095 & 0.1849 ± 0.0068 & 0.4379 ± 0.0076 & 0.4916 ± 0.0083\\
        & MultiVAE & 0.1377 ± 0.0190 & 0.1805 ± 0.0319 & 0.4361 ± 0.0099 & 0.4967 ± 0.0259\\
        & $\beta$-VAE & 0.1053 ± 0.0234 & 0.2010 ± 0.0394 & 0.4287 ± 0.0230 & 0.4645 ± 0.0312\\
        & MacridVAE & {\textbf{0.1593 ± 0.0219}} & 0.2527 ± 0.0230 & {\textbf{0.4782 ± 0.0162}} & {\textbf{0.5611 ± 0.0341}}\\
        \hline
    \end{tabular}
    \label{tab:disen_inter}
\end{table}

In \cref{tab:disen_inter} we show the disentanglement and representation interpretability results of the considered models. We observe that PureSVD has the highest \emph{completeness} across all datasets and the highest \emph{disentanglement} in 2/4 datasets. MacridVAE, despite explicitly aimed at learning disentangled representations in RSs, shows the best \emph{disentanglement} only in GoodReads-Children dataset. In the other datasets, its \emph{disentaglement} and \emph{completeness} scores are much lower than PureSVD, and MultiVAE in ML1M. In~\cite{ma2019learning}, the authors provide only a qualitative inspection of the representation space learned by MacridVAE (figure 2 in~\cite{ma2019learning}). While their results indicate that MacridVAE can encode the items' ground truth category in the latent space, our quantitative evaluations show that its disentanglement can be surpassed by simpler models. Nevertheless, the interpretability of the representations of MacridVAE as measured by LIME-/SHAP-global is the best among our reproduced models. This is in line with the original work findings (figure 3 in~\cite{ma2019learning}) where the authors show that MacridVAE learns human-interpretable concepts in the dimensions of the latent space.

\subsection{Effect of disentanglement on effectiveness (RQ2)}
\begin{figure}
    \centering
    \includegraphics[width=\linewidth]{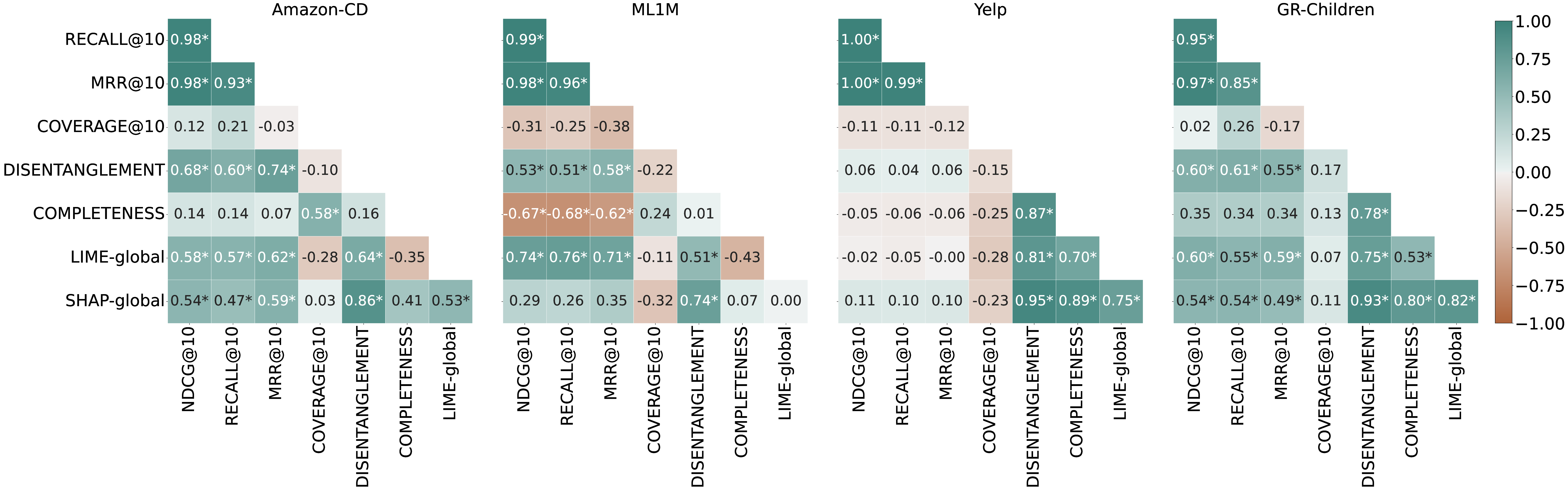}
    \caption{Repeated measurements correlation of effectiveness, disentanglement and representation interpretability measures for each dataset. (*) denotes statistical significance at $\mathbf{p < 0.05}$.}
    \label{fig:rmcorr_datasets}
\end{figure}

\begin{figure}
    \centering
    \includegraphics[width=\linewidth]{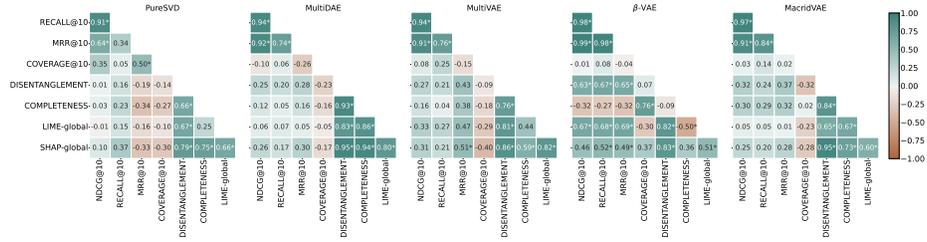}
    \caption{Repeated measurements correlation of effectiveness, disentanglement and representation interpretability measures for each model. (*) denotes statistical significance at $\mathbf{p < 0.05}$.}
    \label{fig:rmcorr_models}
\end{figure}

We study the relation between effectiveness, disentanglement and interpretability of learned representations through statistical correlation. Given that results within a model and/or dataset are correlated and violate the i.i.d assumption of common correlation coefficients, we use \emph{repeated measurements correlation}~\cite{bakdash2017repeated} (RMCORR) that considers intra-group correlations. We present the intra-model RMCORR per dataset in \cref{fig:rmcorr_datasets} and the intra-dataset RMCORR per model in \cref{fig:rmcorr_models}.

In \cref{tab:effectiveness}, MacridVAE shows the best recommendation effectiveness scores while being one of the models with the lowest disentanglement. While Ma et al.~\cite{ma2019learning} attribute MacridVAE's state-of-the-art performance to its ability to disentangled user intentions, our reproducibility study shows a lack of consistent and statistically significant correlation between disentanglement and effectiveness measures; when accounting for the models (\cref{fig:rmcorr_datasets}) we observe a statistically significant correlation, however, no statistically significant correlation is observed when accounting for the datasets (\cref{fig:rmcorr_models}). This lack of correlation can be observed also for the other models and especially evident for PureSVD, which despite its highest \emph{disentanglement} and \emph{completeness} scores, falls behind the neural models in recommendation performance. On the other hand, the lack of correlation between disentanglement and effectiveness does not support the alleged trade-off between disentanglement and downstream task accuracy~\cite{burgess2018understanding,kim2018disentangling,lezama2019overcoming} in the RSs datasets and models that we study.

\subsection{Effect of disentanglement on representation interpretability (RQ3)}
In \cref{fig:rmcorr_datasets,fig:rmcorr_models} we observe a strong positive correlation (RMCORR $\in [0.51, 0.95]$) between disentanglement and LIME-/SHAP-global measures. This correlation holds across all the datasets and models that we run, especially for MacridVAE. This validates Ma et al.~\cite{ma2019learning} claim that MacridVAE is able to encode user intents on the dimensions of the learned latent space. While one of the benefits of learning disentangled representations is better representation interpretability~\cite{bengio2013representation,chen2016infogan,wang2022disentangled,locatello2019challenging,rudin2022interpretable}, to our best knowledge this is the first study to explicitly quantify this connection in RSs.

\subsection{Limitations}
In the context of RSs, the lack of ground truth factors of variation can prevent an objective evaluation of disentanglement. In this reproducibility study, similar to existing literature~\cite{nema2021disentangling,wang2020disentangled,10.1145/3539618.3591665,zhang2024denoising,zheng2021disentangling,zhao2022popularity,yang2023debiased}, we derive factors of variation from a combination of item content information and interaction data, which could have affected our ability to reproduce the disentanglement results of Nema et al.~\cite{nema2021disentangling}.
In this work, we investigate the link between disentanglement and interpretability with feature attribution methods like LIME and SHAP. However, both of these methods have their own limitations~\cite{kaur2020interpreting} which our proposed LIME-/SHAP-global inherit. Finally, we note that we focus here in the interpretability of representations and how they relate to the ground truth factors of variation, which does not necessarily translate in the interpretability of the entire model and its downstream task output~\cite{rudin2022interpretable}.

\section{Conclusion}
We presented the first reproducibility study of representation disentanglement in RSs and its association to recommendation effectiveness and representation interpretability. In our study, we found that it is non-trivial to reproduce disentanglement results without access to the ground truth factors of variation of existing work. The differences between our reproduced recommendation effectiveness scores and those reported in existing work are within 10\% in ML1M and within 43\% in GoodReads-Children. We also presented an adaptation of LIME and SHAP for quantifying representation interpretability. Our correlation analysis on the link between disentanglement and interpretability showed a strong positive correlation between the two, supporting qualitative evidence on their direct connection of prior work~\cite{ma2019learning,wang2020disentangled,zheng2021disentangling,guo2022topicvae,de2023disentangling}. On the other hand, different from existing literature in representation disentanglement in other domains~\cite{higgins2016beta,kim2018disentangling,lezama2019overcoming,rudin2022interpretable}, we did not find a consistent and statistically significant correlation between disentanglement and recommendation effectiveness in the datasets and models that we used.

%
%
%
\bibliographystyle{splncs04}
\bibliography{bib}
\end{document}